\documentclass[a4paper,11pt]{article}
\pdfoutput=1 

\usepackage{jinstpub} 
\usepackage{siunitx}
\usepackage[euler]{textgreek}
\usepackage[colorinlistoftodos,prependcaption]{todonotes}

\title{4H-Silicon Carbide as particle detector for high-intensity ion beams}

\author{Manuel Christanell, }
\author{Maximilian Tomaschek }
\author{and Thomas Bergauer}
\affiliation{Institute of High Energy Physics of the Austrian Academy of Sciences (HEPHY)\\ Nikolsdorfer Gasse 18, 1050 Wien, Vienna, Austria}

\emailAdd{thomas.bergauer@oeaw.ac.at}

\abstract{In ion cancer therapy, high-intensity ion beams are used to treat tumors by taking advantage of the Bragg-Peak. Typical ion therapy centers use particle rates up to $10^{10}$ ions/second for treatment. On the other hand, such intensities are often too high when using these beamlines for particle physics experiments or as a test-beam environment in general. The project presented here aims to develop a beam position and intensity monitor, to cover a wide intensity range from a few Hz up to GHz rates, as used in clinical settings.

Silicon carbide (SiC) is an attractive detector material for this application because it combines potential high radiation hardness with high thermal conductivity to avoid cooling. Moreover, its high electron saturation velocity allows very fast signals to mitigate pile-ups. However, some special properties of the material like different crystal polytypes have to be considered.

In this paper, measurements on both a pad and a micro-strip SiC sensor prototype of 4H lattice geometry are shown. The sensors were tested in the laboratory using radioactive sources and with a proton beam in a wide intensity range (kHz-GHz) and with different energies (60-800 MeV) available at MedAustron, an ion cancer therapy center located in Austria.

The measurements show that MIP particles cannot be detected reliably with the used discrete electronics setup in combination with the single-channel sensor. However, the strip sensor combined with an ASIC-based readout electronics from the CMS/Belle-II experiments allows to recover a certain part of the signal. This makes it possible to determine the ionization energy and average number of electron/hole pairs generated in the studied sensor samples.
}


\proceeding{22$^{\text{nd}}$ International Workshop on Radiation Imaging Detectors\\
  27 June 2021 to 1 July 2021\\
  Online}

\begin{document}
\maketitle
\flushbottom

\section{Introduction}
Monitoring of high-intensity ion beams requires sophisticated detector systems. These are used, among others, in particle therapies for cancer treatment, where particle rates up to $10^{10}$ ions/s, concentrated on a small area of approx. 8$\times$8\;mm$^2$, occur~\cite{lowflux}. Special requirements apply here also on the corresponding sensor materials. These cover a high radiation hardness, which limits the wear of the material, affordability to manufacture large sensor areas, and high charge carrier velocities to reduce the chance of pile-ups.

A promising and relatively new sensor material in the field of particle detection is silicon carbide (SiC). It has some benefits over other materials like silicon. The higher band-gap decreases the dark current and eliminates the need for cooling even in harsh environments \cite{Ruddy2003}. The potential higher radiation hardness leads to fewer material alterations, resulting in a longer lifetime, especially under high radiation conditions \cite{Ruddy2003}. Also, a temperature-independent performance up to a few $\SI{100}{\degreeCelsius}$ was observed \cite{Metzger2002}. The characteristics of SiC make the material also interesting for power devices in the chip industry. This increases the manufacturing volume and, in turn, reduces production costs to make large area detectors available and affordable. This is in contrast to diamond sensors produced in the chemical vapor deposition (CVD) procedure, which are still bounded to a few $\SI{}{\square\cm}$.



\section{Silicon Carbide for Particle Detectors}
\subsection{General Properties}
A comparison of selected parameters for Si and SiC are shown in table \ref{tab:electrical_properties}. These parameters show higher ionization energies, band-gap and saturated electron velocities for SiC compared to Si. The charge mobility of SiC is lower compared to the one of Si. However, due to the combination of high electric breakdown field and high drift velocity, the signal generated in SiC is faster than the one in Si.

There are two commonly produced SiC polytypes, one being cubic (3C-SiC) and two hexagonal (4H-SiC and 6H-SiC). The main difference between these different cell structures is the anisotropy effects, which occur for the hexagonal primitive cells. The anisotropy of the hexagonal SiC polytypes show a strong variation in the electron mobility depending on the lattice axis $c$. 4H-SiC is more commonly used as it has a higher electron mobility and the anisotropic effect is not as high compared to 6H-SiC.

\begin{table}[htb]
\caption{Electrical properties of SiC, Si and diamond. These properties are taken from \cite{harris1995properties, hara1998silicon, sze2008semiconductor} if not indicated otherwise. MPV stands for most probable value.}
\label{tab:electrical_properties}
\centering
\begin{tabular}{|l|l|l|l|l|}
\hline
& \textbf{Si} & \textbf{4H-SiC} & \textbf{6H-SiC} & \textbf{3C-SiC} \\ \hline
\textbf{Bandgap Energy} $E_{\text{g}}$ {[}eV{]} & 1.12 &    3.26    &   3.03     &  2.4      \\ \hline
\begin{tabular}[c]{@{}l@{}}\textbf{Ionization Energy} $E_{\text{i}}$ {[}eV{]}\\ generating one e/h-pair\end{tabular}&  3.64  &  \begin{tabular}[c]{@{}l@{}}5 \cite{4H_SiC_5_ionization}
, 8.6 \cite{4H_SiC_8_6_ionization}\end{tabular}      &    9 \cite{6H_SiC_9_ionization}    &   -     \\ \hline
\begin{tabular}[c]{@{}l@{}}\textbf{Breakdown Field} $E_{\text{B}}$ [MV/cm]\\ for $N_{\text{D}}$ = 10$^{17}$\,cm$^{-3}$\end{tabular} &  0.3  &    \begin{tabular}[c]{@{}l@{}}$\perp$ c: -\\ $\parallel$ c: 3.0\end{tabular}    &  \begin{tabular}[c]{@{}l@{}}$\perp$ c: 3.2\\ $\parallel$ c: > 1\end{tabular}       &    > 1.5    \\ \hline
\begin{tabular}[c]{@{}l@{}}\textbf{Electron Mobility} \\ $\mu_{\text{e}}$ [cm$^{2}$/Vs]\end{tabular}& 1430   &   \begin{tabular}[c]{@{}l@{}}$\perp$ c: 800\\ $\parallel$ c: 900\end{tabular}     &    \begin{tabular}[c]{@{}l@{}}$\perp$ c: 400\\ $\parallel$ c: 60\end{tabular}    &   800     \\ \hline
\begin{tabular}[c]{@{}l@{}}\textbf{Hole Mobility} \\ $\mu_{\text{h}}$ [cm$^{2}$/Vs]\end{tabular}& 480   &     115   &     90   &  40      \\ \hline
\begin{tabular}[c]{@{}l@{}}\textbf{Saturated Electron}
\textbf{Velocity}\\ v [10$^{7}$cm/s]\end{tabular}&  1  &   2.2 \cite{SiC_sat_velo_2_2}    &    2    &      2.5  \\ \hline
\begin{tabular}[c]{@{}l@{}}\textbf{e-h-pair/μm generation}
\\ by one MIP\end{tabular}& \begin{tabular}[c]{@{}l@{}}MPV: 72\\ mean: 106\end{tabular} &  MPV: 55  \cite{SiC_Si_MIP_e_h_gneration} &    -    &      -  \\ \hline
\end{tabular}
\end{table}

\subsection{Samples used in this Study}
The SiC sensor samples used for this study were fabricated by CNM Barcelona based on 4H-SiC wafers with a diameter of 4~inches, which were purchased from Ascatron\footnote{Ascatron Sweden, Webpage: \url{http://ascatron.com}}. The wafers contain a 50\,μm thick, n-doped epitaxial active layer with a resistivity of 20\,Ohm\,cm. The wafers' total thickness is 400\,μm.
A schematic cross section showing the layer composition of the device is shown in~\cite{rafiElectronNeutronProton2020} and more details on the contact window is given in~\cite{rafi10MthickFourquadrant2017}.

The wafer layout contains multiple structures, arranged in a grid structure, from which planar pad diodes and micro-strip sensors were used in this publication.
The outer dimensions of the pad sensor are 3x3\,mm$^2$, including guard rings and periphery.
The strip sensor consists of 64 strips with a pitch of 50\,μm and a strip length of 3\,mm.
Both the pad and the strip sensors were fixed by conductive glue to mechanical supports providing the backside HV contact. The front-side contact(s) to the readout electronics were made by wire-bonding (25µm thick Al-1\%Si wire) to external contact pads on the UCSC board PCB (for the pad sensor) and to an Aluminium-on-glass pitch adapter (for the strip sensor), respectively.

\section{Setups}

\subsection{Pad sensors with UCSC Boards}

\begin{figure}[htb!]
    \centering
    \begin{minipage}[t]{0.396\linewidth}
    \includegraphics[width=1\linewidth]{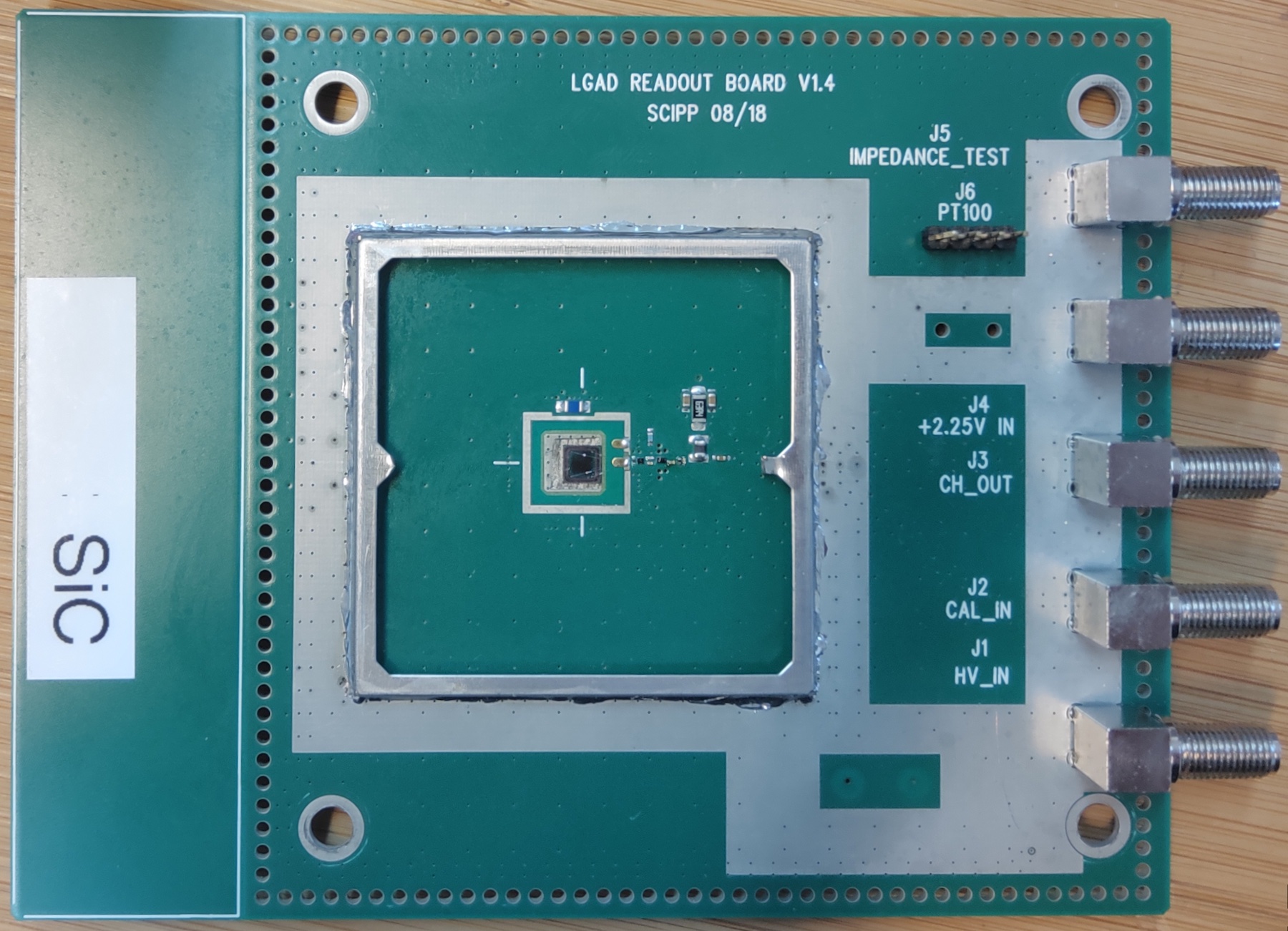}
    \caption{The silicon carbide single-channel sensor bonded on the UCSC (LGAD) readout board V1.4.}
    \label{fig:SiC_singleChannel_UCSC}
    \end{minipage}
    \hfill
    \begin{minipage}[t]{0.59\linewidth}
    \centering
    \includegraphics[width=1\linewidth]{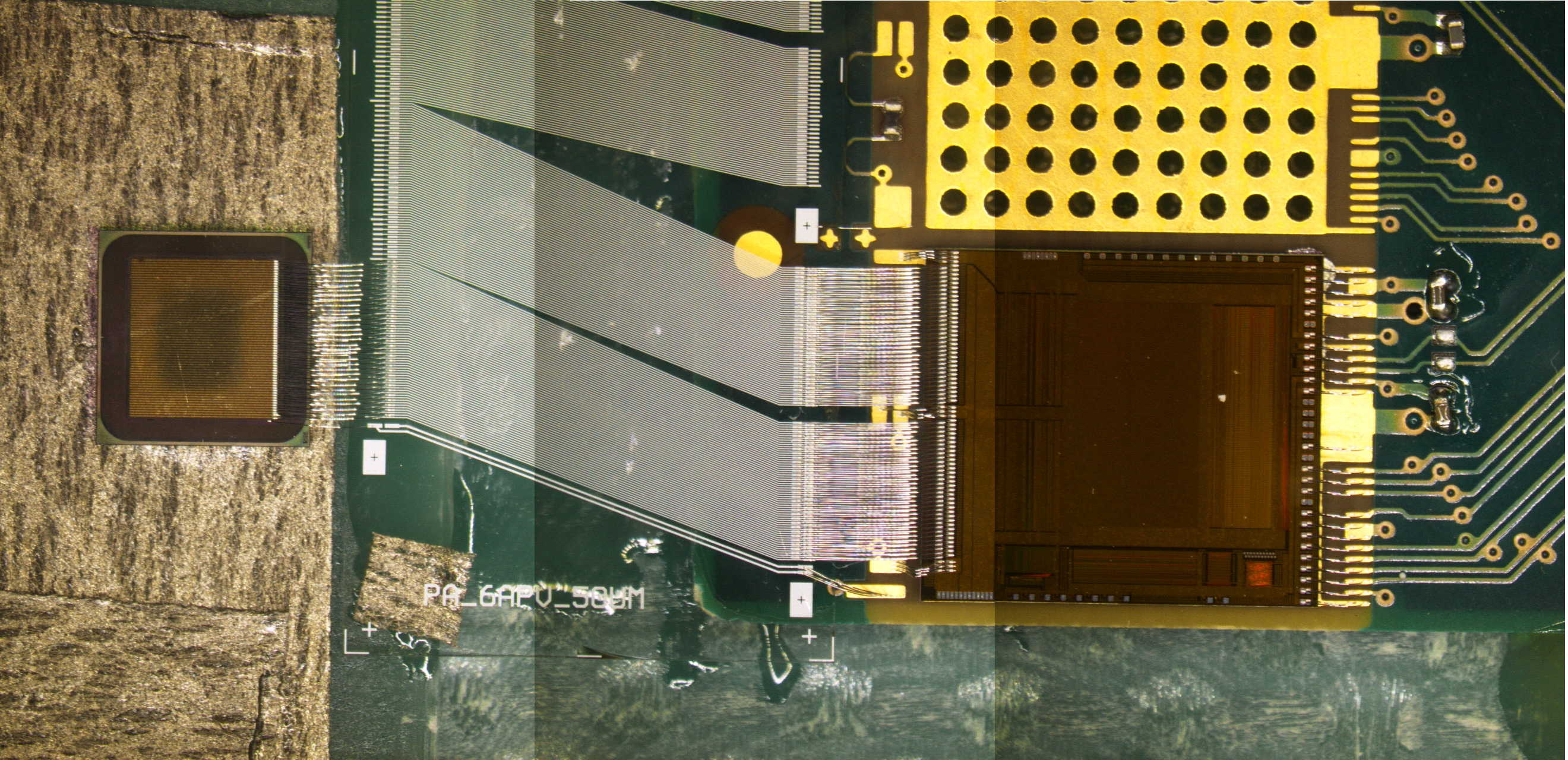}
    \caption{SiC strip sensor (left) connected by wire-bonds and an intermediate pitch-adapter (middle) to an APV25 readout chip mounted on a PCB board (right).}
    \label{fig:SiC_strip_sensor_APV25_hybrid}
    \end{minipage}
\end{figure}

The single-channel silicon carbide sensor was operated with the UCSC readout board V1.4 (figure \ref{fig:SiC_singleChannel_UCSC}), developed by the University of California Santa Cruz (UCSC) and widely used for single-channel sensor characterization measurements. The board is often used with LGAD sensors and therefore also known as {\it LGAD board}. The central part of the board is a single-stage inverting transimpedance amplifier. According to the developers, the transimpedance equals $\SI{470}{\ohm}$ with a bandwidth of $\SI{1.6}{\GHz}$. In this study, the original feedback resistor R14 of $\SI{470}{\ohm}$ was replaced by a resistor of $\SI{2.2}{\kohm}$, to increase the amplification of the circuit.

\begin{figure}[htb]
    \centering
    \includegraphics[]{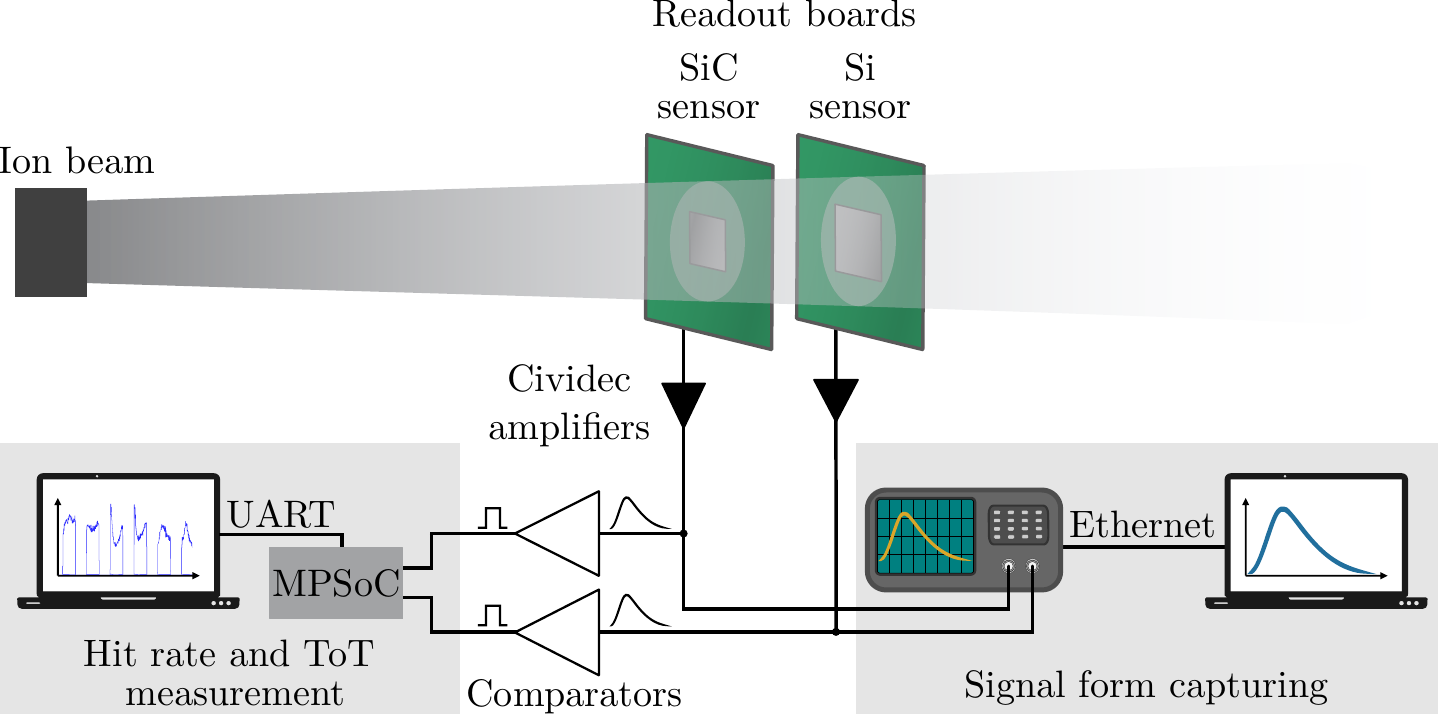}
    \caption{Measurement setup with single-channel UCSC boards equipped with Si and SiC sensors, as used in the beam test.}
    \label{fig:MedAustron_setup_singleChannel}
\end{figure}

At MedAustron, the single-channel silicon carbide (upstream) and a planar silicon diode (downstream) were operated on separate UCSC boards in a proton beam (figure \ref{fig:MedAustron_setup_singleChannel}) with an energy between 62 and 800˜MeV. 
The bias voltages were kept constant at $\SI{400}{\V}$, and the output signals were amplified by the Cividec "C2-HV" amplifier (Bandwidth: $\SI{2}{\GHz}$, Gain: $\SI{40}{\dB}$) before being further processed. The signal from the silicon diode was used to trigger the capturing of the SiC waveform by a Tektronix DSA 70804 Oscilloscope (Bandwidth: $\SI{8}{\GHz}$) with the FastFrame mode. At the same time, both signals were first discriminated by two comparators and then counted by the FPGA included in the "Xilinx Zynq UltraScale+" MPSoC. Additionally, the FPGA recorded the total time-over-threshold (ToT). Both values together with a local timestamp ($\SI{1.5}{\GHz}$ clock count) were then transferred over UART to a PC, where the whole data was stored and the hit rates displayed~\cite{MChristanell}.

The lab measurements with an Americium-241 alpha source included only the silicon carbide detector. In contrast to the MedAustron setup, no second amplification, no discrimination, and particle rate detection were done. In addition, the sensor bias voltage was varied between $\SI{100}{\V}$-$\SI{400}{\V}$. The Americium source was placed above the detector. When a particle hit the sensor, the produced UCSC output signal was fed into the Tektronix oscilloscope. If the signal was above the configured trigger level, its waveform was captured. After capturing a certain amount of hits, the acquired wave forms were transferred and saved to a PC.


\subsection{Strip Sensor Readout with APV25 ASIC}

The readout of the SiC strip sensor was utilized using an APV25 readout ASIC connected to the sensor on a hybrid board (see figure \ref{fig:SiC_strip_sensor_APV25_hybrid})~\cite{Hirishima_paper}. The APV25 is read out by a FADC system which was originally developed for the Belle-II project~\cite{THALMEIER2017633}.
The APV25 provides 128 channels for strip sensor readout and includes a preamplifier, pulse shaper and integrating circuit. Afterwards the analog signal is converted and processed by a FADC board.  The data is read out by the EPICS-based run- and slow-control \cite{Hirishima_paper}.

At MedAustron the setup consist of the SiC sensor module as DUT and a beam telescope comprising four Si tracking planes, originally built for developing an ion imaging modality~\cite{Hirishima_paper}. 
The four tracker modules are equipped with double-sided strip detectors (DSSD). The material of the sensors is n-type silicon and the
nominal thickness of the sensors is 300\,μm. The active area (25×50\,mm\textsuperscript{2}) feature 512 AC-coupled strips on each side, with a pitch of 50\,μm on the y-axis (p-side) and 100\,μm on the x-axis (n-side) \cite{Hirishima_paper}.

Acquired raw data can be processed by three parameters to set a threshold for noise and signal. The threshold is defined by a seed cut parameter multiplied by the width of the Gaussian distributed noise.
The other parameters define how many consecutive samples digitized by the APV25 within a 25\,ns sample rate are required over the set threshold to count as an event (called MinHitLength) and how many clusters are allowed per sensor and per event.

\section{Laboratory Measurements}

\begin{figure}[!htb]
    \centering
    \includegraphics[]{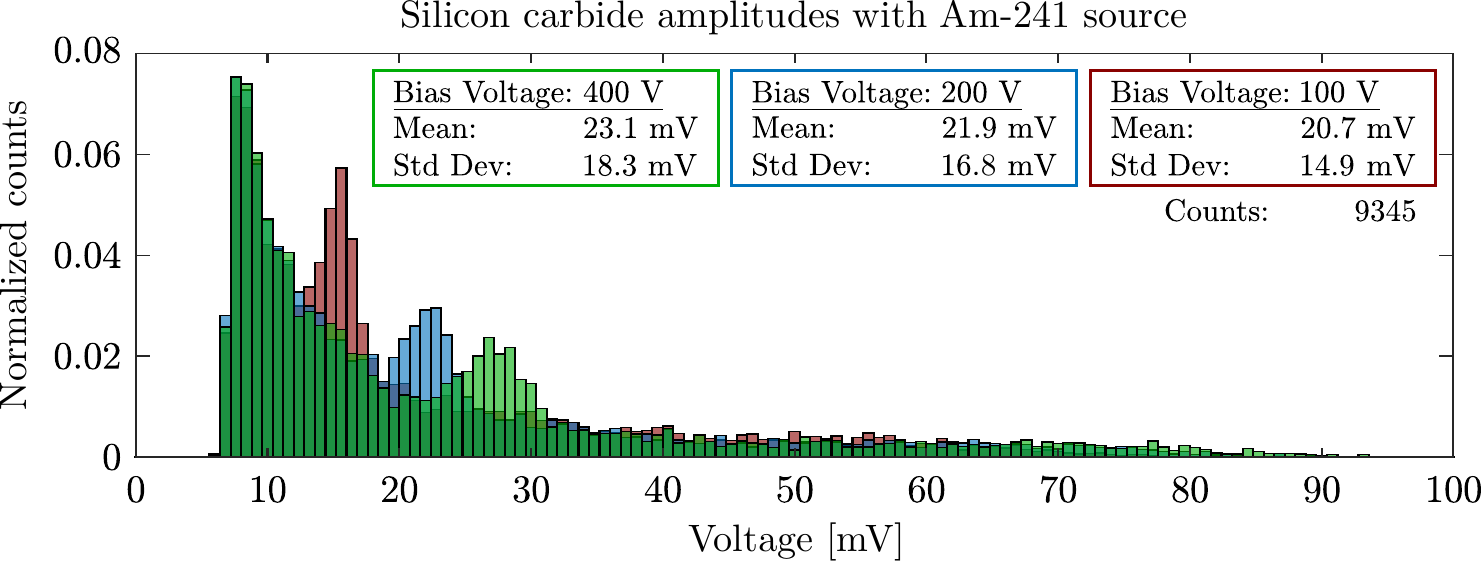}
    \caption{Signal distributions of the single-channel SiC sensor produced by alpha particles at different bias voltages ($\SI{100}{\V}$, $\SI{200}{\V}$, $\SI{400}{\V}$). The colors of the bins were made slightly transparent to make also the covered histogram parts visible.}
    \label{fig:Alpha_result_singleChannel}
\end{figure}

Alpha particles from an Am-241 source have been used to generate signals.
Due to scattering processes in $\SI{4}{\mm}$ air between the Am-241 source and the detector (given by the geometry of the setup) as well as the scattering processes in the "dead layer" of the sensor entrance window \cite{JRaf}, the discrete peaks in the spectrum are widened and produce the measured signal amplitude distribution of figure \ref{fig:Alpha_result_singleChannel}.
The effect of the air passage was verified by Geant4 simulations.

\section{Tests with Particle Beams at MedAustron}
MedAustron is a research and cancer treatment facility located in Wiener Neustadt (Austria). The central part is a synchrotron, producing proton beams with $\SI{62}{\MeV}$-$\SI{250}{\MeV}$ in clinical mode and up to $\SI{800}{\MeV}$ for research purposes. These particle energies $\SI{62}{\MeV}$-$\SI{800}{\MeV}$ correspond to a MIP charge generation equivalent of 5.03-1.14. Besides the nominal high beam intensities ($10^{10}$ particles per $\SI{5}{\s}$ spill), special low flux settings with particle rates around $\SI{3}{\kHz}$, $\SI{400}{\kHz}$ and $\SI{4}{\MHz}$ were used~\cite{lowflux}.
\subsection{Single pad sensors}
\begin{figure}[htb!]
    \centering
    \includegraphics[]{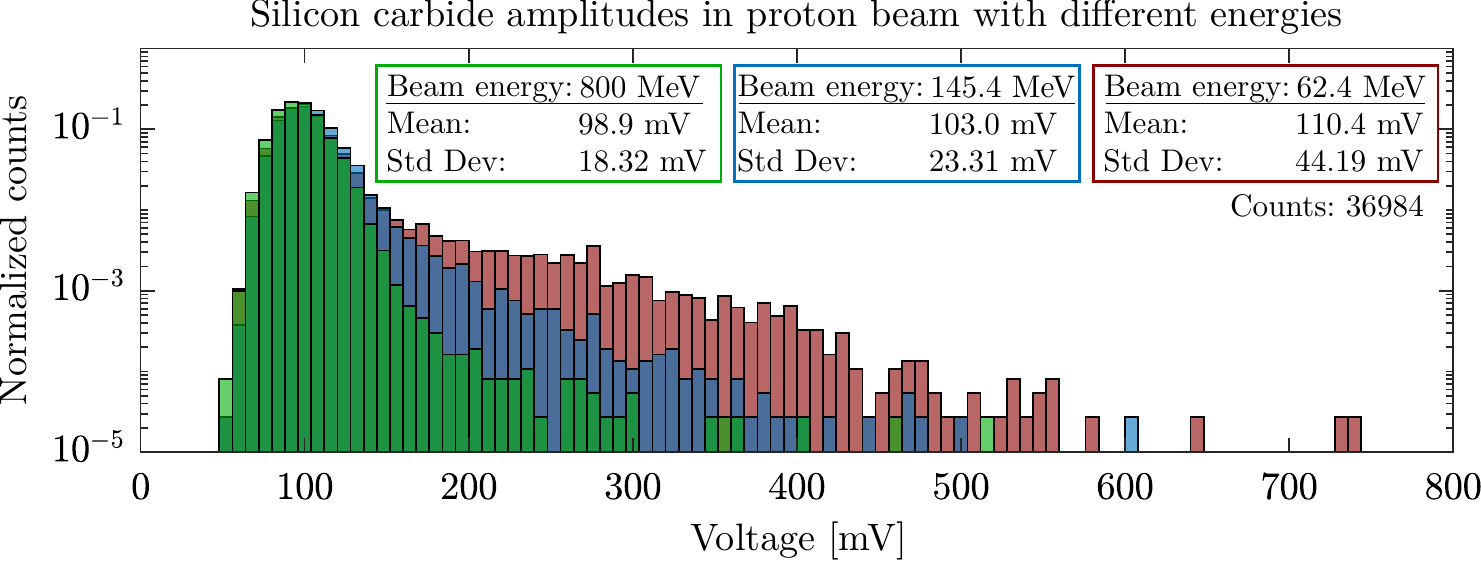}
    \caption{SiC sensor amplitude distributions measured in the beamline
            for different beam energies. The wave forms recording was triggered
            by the silicon sensor. The colors of the bins were made slightly transparent to make also the covered histogram parts visible.}
    \label{fig:MedAustron_result_singleChannel}
\end{figure}

The silicon carbide sensor amplitude histograms, triggered by the silicon sensor, almost show a Gaussian distribution (figure \ref{fig:MedAustron_result_singleChannel}). 
The distribution is a consequence of Gaussian noise, but in contrast to the noise measurement, a Landau tail is visible here. The tail is most pronounced for low beam energies, resulting from higher energy depositions at lower particle energies. The fusion of the Landau and the Gaussian distribution, like seen here, makes it impossible to set a threshold with which all particle hits can be detected reliably. 

\subsection{Strip Sensor}
\begin{figure}[htb!]
    \centering
    \begin{minipage}[t]{0.49\linewidth}
    \includegraphics[width=1\linewidth]{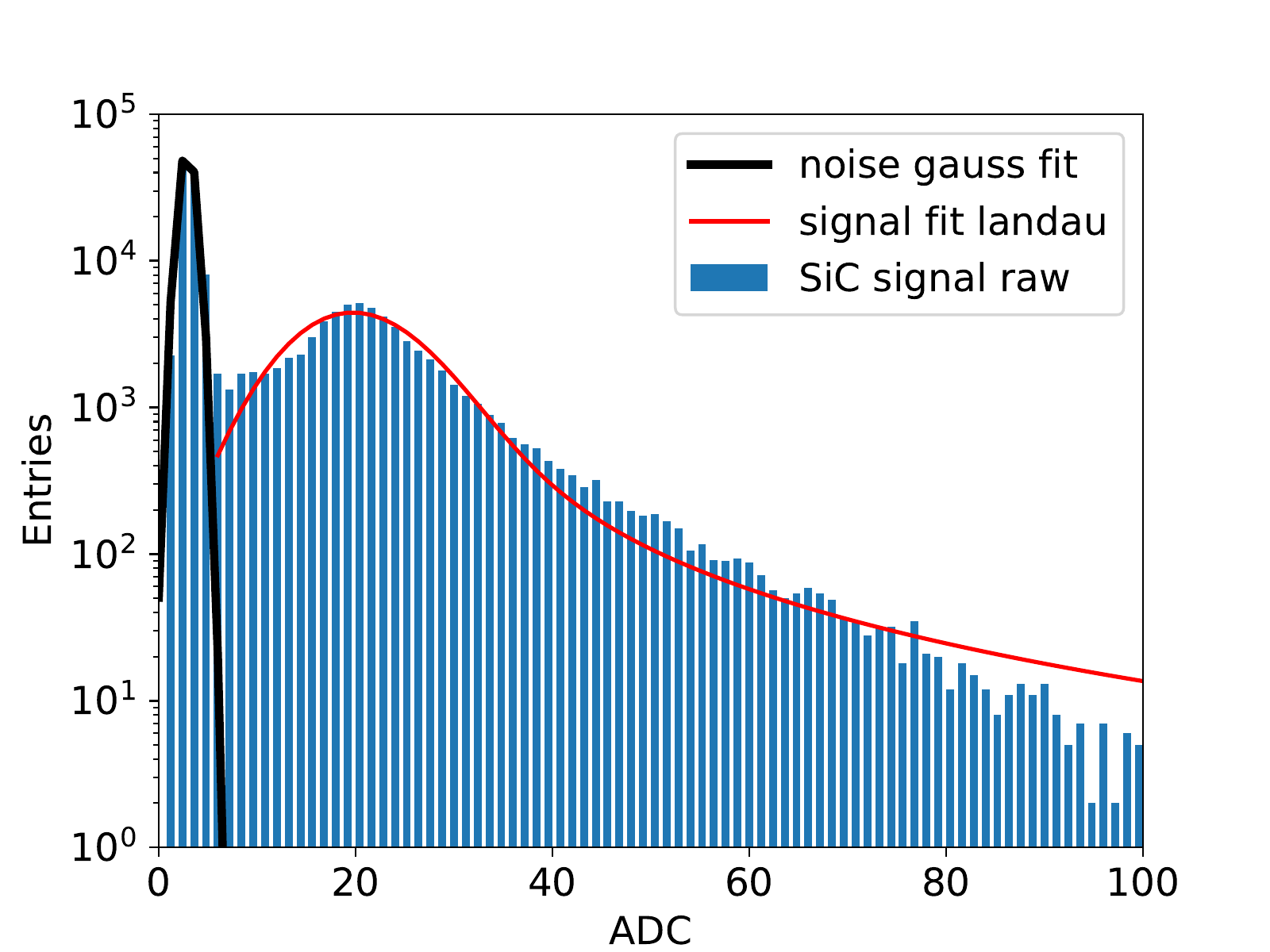}
    \caption{This histogram shows the signal protons with 145\,MeV. On the left the Gaussian shaped noise distribution is visible and on the right the Landau distributed signal is shown.}
    \label{fig:145MeV_gauss_langau_fit_log}
    \end{minipage}
    \hfill
    \begin{minipage}[t]{0.49\linewidth}
    \centering
    \includegraphics[width=1\linewidth]{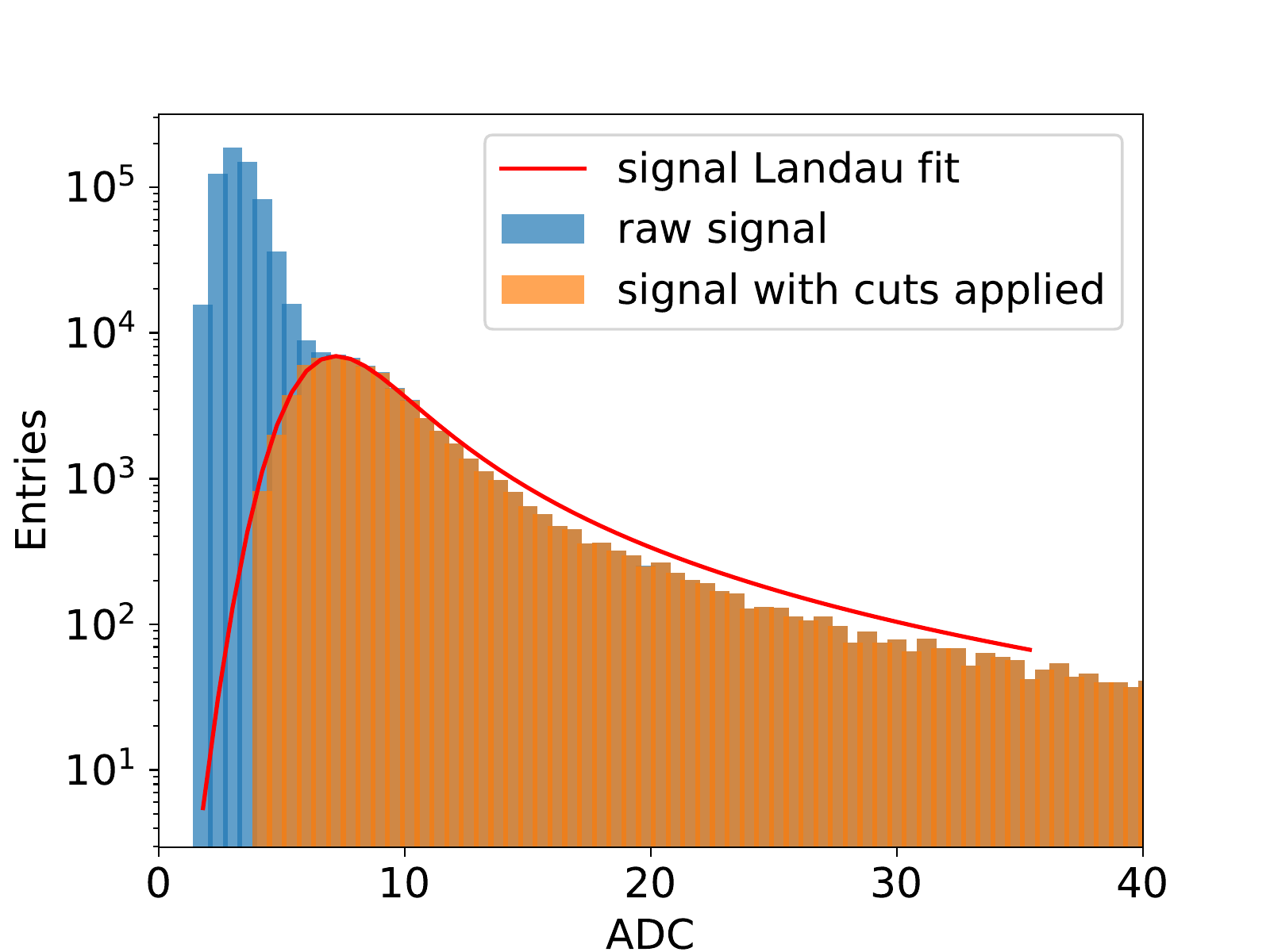}
    \caption{The signal of protons with 800\,MeV overlaps with the noise, so  that a separate signal peak cannot be detected (blue bars). However by applying cuts, the signal gets visible (orange bars).}
    \label{fig:800MeV_raw_cut_log}
    \end{minipage}
\end{figure}

The SiC strip sensors were exposed to protons with an energy of 145\,MeV. A typical signal plot can be seen in figure \ref{fig:145MeV_gauss_langau_fit_log}. This graphs shows the typical Gaussian shaped noise and the Landau distributed signal. The signal and noise is clearly separable for protons with an energy up to 252\,MeV. The signal generated by the traversing particle becomes smaller the higher the energy of the particle beam is. The signal of protons with 800\,MeV overlaps with the noise (see figure \ref{fig:800MeV_raw_cut_log}).

However it is possible to separate the signal from the noise by using the method described above. It was found from the measurement with protons at 145\,MeV that the noise is cut out setting the seed cut parameter above 4.5. This parameter was set up for the 800\,MeV data analysis and at least three samples over threshold were required (MinHitLength=3). With this configuration it was possible to separate the signal from the noise (see figure \ref{fig:800MeV_raw_cut_log}).

\begin{figure}[htb!]
    \centering
    \begin{minipage}[t]{0.49\linewidth}
    \includegraphics[width=1\linewidth]{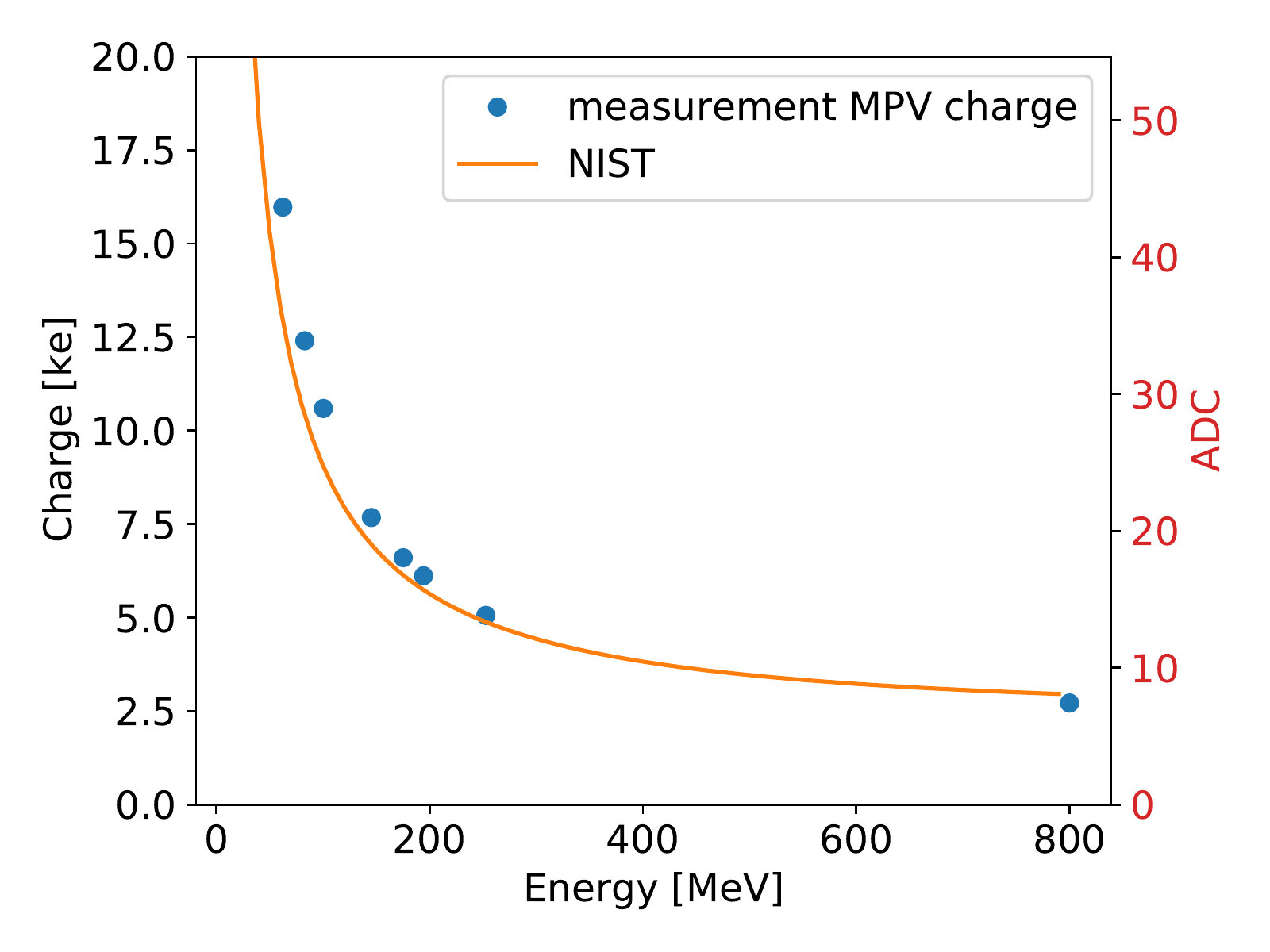}
    \caption{SiC strip sensor energy scan (protons with an energy ranging from 62 MeV up to 800 MeV) with a bias voltage of 500 V applied.}
    \label{fig:energy_scan_converted}
    \end{minipage}
    \hfill
    \begin{minipage}[t]{0.49\linewidth}
    \centering
    \includegraphics[width=1\linewidth]{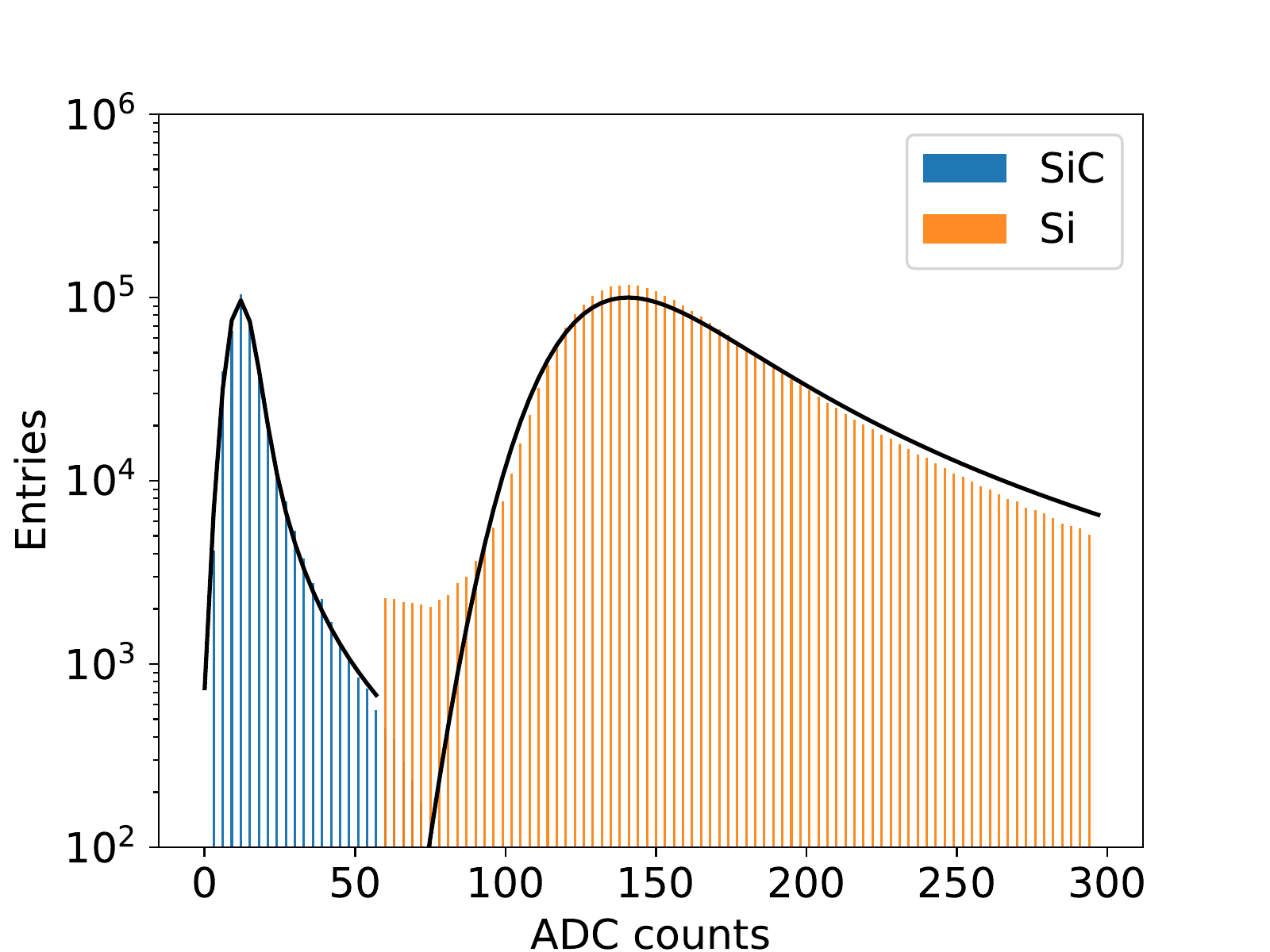}
    \caption{Signal comparison of the SiC and Si detector with a proton energy of 252\,MeV. The MPV of the Landau distributions were found at 14.76 ADC and 142.3 ADC respectively.}
    \label{fig:SiC_Si_signal_comparison}
    \end{minipage}
\end{figure}

An energy scan (see figure \ref{fig:energy_scan_converted}) was conducted to compare the measured results to the Bethe-Bloch equation, using the NIST PSTAR data base \cite{nist}. The proton energies are to low to show the energy for a minimum ionizing particle (MIP). An Allpix\textsuperscript{2} simulation indicates that the MIP energy is reached at around 5\,GeV~\cite{MTomaschek}.

The ionization energy value of 4H-SiC is given in the literature between 5\,eV to 8.6\,eV \cite{4H_SiC_5_ionization, 4H_SiC_8_6_ionization}. The ionization energy can be calculated by comparing the measurements of the SiC sensor (active thickness of 50\,μm) and a Si tracker sensor (active thickness of 300\,μm) . By comparing the signal heights of the different material (see figure \ref{fig:SiC_Si_signal_comparison}) normalized to the sensor's active thickness and acknowledging that Si has a ionization energy of 3.64\,eV the ionization energy of 4H-SiC can be calculated as 5.85\,eV \cite{MTomaschek}.

Normalizing the collected charge in the sensor to the active thickness of the sensor and the corresponding MIP equivalent (charge normalized to the charge of a MIP) value gives the number eh-pairs generated by one MIP. For the MPV of the signals for the different energies a value of eh-pairs was found at 57.1$\pm$3.9 eh-pairs/μm \cite{MTomaschek}. This eh-pair generation number fits well with the given value in literature of 55 eh-pair/μm \cite{SiC_Si_MIP_e_h_gneration} for a MIP particle.

\section{Conclusion}
The study shows that MIP particles cannot be detected reliably with the used discrete electronics setup consisting of UCSC boards combined with single-channel sensors. However, the strip sensor operated with an ASIC-based readout electronics from the CMS/Belle-II experiments allows recovering a certain part of the signal by applying special cuts sensitive to the signal form. This makes it possible to determine the ionization energy and the average number of electron/hole pairs generated in the studied sensor samples.


\acknowledgments

We gratefully received the SiC samples used for this publication from IMB-CNM-CSIC Barcelona.
Parts of this work was possible by funding from the Austrian research promotion agency FFG, project number 883652.


	\bibliography{SiC-paper_iWorid}

\end{document}